\documentclass[twocolumn,aps,prd,preprintnumbers,superscriptaddress,nofootinbib,10pt]{revtex4-1}
\usepackage{epsfig}
\usepackage{graphicx}
\usepackage{psfrag}
\usepackage{amsmath,amssymb}
\usepackage{colordvi}
\usepackage{amsfonts}
\usepackage{enumerate}
\usepackage{slashed}
\usepackage{color}
\usepackage{xcolor}

\begin{document}
\title{Perturbative QCD Analysis of Near Threshold Heavy Quarkonium Photoproduction at Large Momentum Transfer }

\author{Peng Sun}
\affiliation{Department of Physics and Institute of Theoretical Physics,
Nanjing Normal University, Nanjing, Jiangsu 210023, China}

\author{Xuan-Bo Tong}
\affiliation{ CAS Key Laboratory of Theoretical Physics, Institute of Theoretical Physics, Chinese Academy of Sciences, Beijing 100190, China}
\affiliation{School of Physical Sciences,University of Chinese Academy of Sciences,Beijing 100049,China}

\author{Feng Yuan}
\affiliation{Nuclear Science Division, Lawrence Berkeley National
Laboratory, Berkeley, CA 94720, USA}

\begin{abstract}
We apply perturbative QCD to investigate the near threshold heavy quarkonium photoproduction at large momentum transfer. From an explicit calculation, we show that the conventional power counting method will be modified and the three quark Fock state with nonzero orbital angular momentum dominates the near threshold production. It  carries a power behavior of $1/(-t)^5$ for the differential cross section. We further comment on the impact of our results on the interpretation of the experiment measurement in terms of the gluonic gravitational form factors of the proton.
\end{abstract}
\maketitle

\section{Introduction}

In Refs.~\cite{Kharzeev:1995ij,Kharzeev:1998bz}, Kharzeev and collaborators proposed the near threshold photoproduction of heavy quarkonium as a way to measure the trace anomaly contribution to the proton mass~\cite{Shifman:1978zn,Ji:1994av,Ji:1995sv,Hatta:2018sqd,Metz:2020vxd,Hatta:2020iin,Ji:2021pys,Ji:2021mtz}. It has attracted a strong interest from the community~\cite{Gryniuk:2016mpk,Hatta:2018ina,Ali:2019lzf,Hatta:2019lxo,Boussarie:2020vmu,Mamo:2019mka,Gryniuk:2020mlh,Wang:2019mza,Zeng:2020coc,Du:2020bqj,Kharzeev:2021qkd,Wang:2021dis,Hatta:2021can,Mamo:2021krl,Kou:2021bez,Guo:2021ibg,Strakovsky:2019bev,Strakovsky:2020uqs,Pentchev:2020kao} due to potential measurements of these processes at the current and future facilities, including JLab-12GeV~\cite{Dudek:2012vr,Chen:2014psa}, electron-ion colliders (EIC) in US~\cite{Accardi:2012qut,AbdulKhalek:2021gbh} and China~\cite{Anderle:2021wcy}. The ultimate goal of these studies is to identify the origin of the proton mass~\cite{Joosten:2018gyo}.  

The original arguments of Refs.~\cite{Kharzeev:1995ij,Kharzeev:1998bz} are based on the vector-meson-dominance and the expansion near the threshold $J/\psi N\to J/\psi N$ system~\cite{Peskin:1979va,Bhanot:1979vb,Luke:1992tm}. Progress has been made to compute directly the differential cross section for $\gamma N\to J/\psi N$ in various models and more recently in QCD analysis~\cite{Boussarie:2020vmu,Hatta:2021can,Guo:2021ibg}. These developments are greatly needed to build a solid ground for the future measurements. The goal of this paper is to show how we can apply perturbative QCD to understand the near threshold heavy quarkonium production.

Near the threshold region, the momentum transfer is large: $-t\sim 2{\rm GeV}^2$ and $10{\rm GeV}^2$ for $J/\psi$ and $\Upsilon$, respectively, where $t$ is the momentum transfer squared from the nucleon target. The large momentum transfer makes a strong argument to apply perturbative QCD. The large $(-t)$ behavior can be calculated following the factorization of nucleon form factor calculations~\cite{Lepage:1979za,Lepage:1980fj,Brodsky:1981kj,Efremov:1979qk,Chernyak:1977as,Chernyak:1980dj,Chernyak:1983ej,Belitsky:2002kj,Tong:2021ctu} and the non-relativistic QCD (NRQCD)~\cite{Bodwin:1994jh} for the heavy quarkonium production. The differential cross section will depend on the associated distribution amplitudes of the nucleon and the NRQCD matrix element of heavy quarkonium. 

An immediate outcome of our analysis is the power behavior of the differential cross section at large $(-t)$. The power behavior has been assumed in the phenomenological studies, see, e.g., Refs.~\cite{Frankfurt:2002ka,Ali:2019lzf,Kharzeev:2021qkd,Wang:2021dis}. Our calculations will provide a solid foundation for this practice. We will also show, more importantly, the conventional power counting method~\cite{Brodsky:1973kr,Matveev:1973ra,Ji:2003fw} have to be modified around threshold. 

We take the threshold limit in our derivations, i.e., $W_{\gamma p}\sim M_V+M_p$, where $W_{\gamma p}$ represents the center of mass energy and $M_V$ and $M_p$ for the heavy quarkonium and proton masses, respectively. To determine the leading contribution, we introduce a parameter~\cite{Brodsky:2000zc}: $\chi=\frac{M_{V}^2+2M_pM_{V}}{W_{\gamma p}^2-M_p^2}$, which goes to 1 at the threshold. We will expand the amplitude in terms of $(1-\chi)$. By applying this expansion, in particular, we will show that the commonly assumed $1/(-t)^4$ power behavior for the differential cross section is actually suppressed by $(1-\chi)$. 

To further simplify our analysis, we apply the heavy quark mass limit with the following hierarchy in scales: $W_{\gamma p}^2\sim M_V^2\gg (-t)\gg\Lambda_{QCD}^2$,
where $\Lambda_{QCD}$ represent the non-perturbative scale. Under this limit, the scattering amplitude can be separated into two parts: the part associated with the photon-quarkonium transition and the part describing gluon interactions with the nucleon states. As a result, the dominant $t$-dependence comes from the nucleon side and can be calculated following that of the gluonic form factors calculations~\cite{Tong:2021ctu}. 

Our framework provides a unique method to unveil the physics mechanism for the threshold heavy quarkonium production. Much of scattering amplitude is calculable in perturbative QCD and can offer an important guidance to build a rigorous formalism for the threshold production process. As mentioned above, the near threshold heavy quarkonium production is dominated by large momentum transfer. That means the power behavior derived in this paper can be applied to most of the experimental data. In particular, we will compare our predictions to recent experimental data from the GlueX collaboration~\cite{Ali:2019lzf}, where the agreement provides a strong indication that perturbative QCD is applicable here. This shall encourage further developments. 

The rest of the paper is organized as follows. We will first examine the threshold kinematics and derive the power counting analysis in Sec.~II. We will focus on the major results from our calculations and discuss the interpretation of these results. We leave the detailed derivations in a separate publication. In Sec.~III, we present phenomenological studies and apply our analysis to recent GlueX data on near threshold $J/\psi$ production at JLab. Finally, we summarize our paper in Sec.~IV.

\section{Near Threshold Kinematics and Power Counting Analysis}

\begin{figure}[tpb]
\includegraphics[width=0.95\columnwidth]{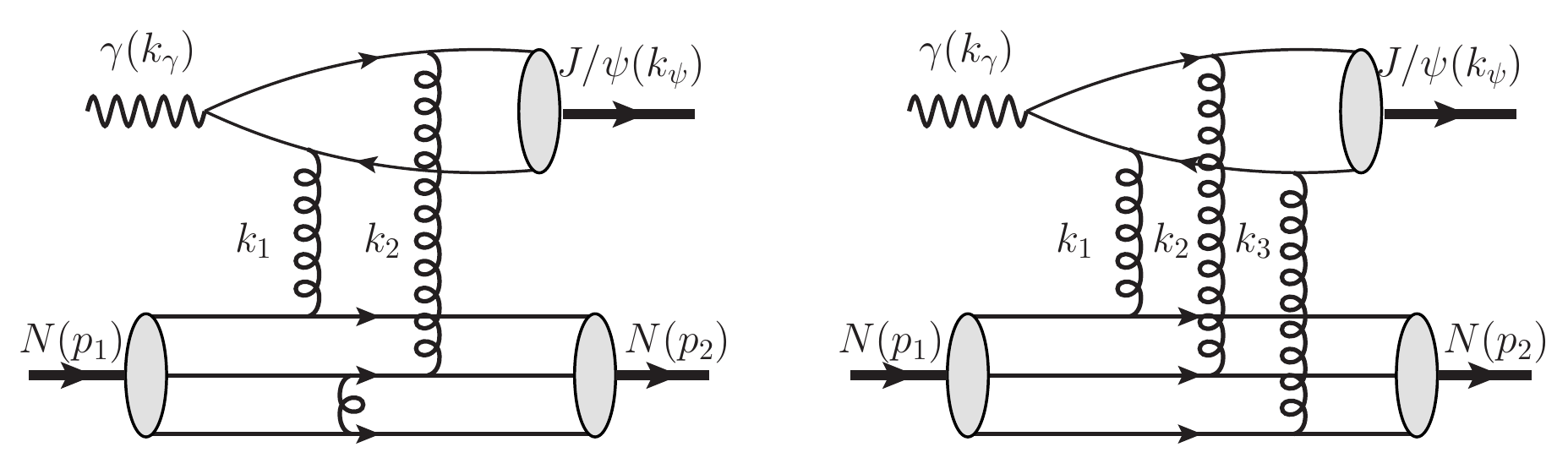}
    \caption{Typical Feynman diagram contributions to threshold $J/\psi$ photoproduction at large momentum transfer with two-gluon exchange (left) and three-gluon exchange (right). The complete results comes from all permutations of the gluon attachments to the upper and lower parts of the above diagrams. Due to $C$-parity conservation, there is no contribution from the three-gluon exchange diagrams.}
    \label{largetdiagram}
\end{figure}

Near threshold heavy quarkonium production is generated through a hard exclusive process with gluon exchange between the heavy quark loop and the nucleon states, as shown in Fig.~\ref{largetdiagram},
\begin{equation}
    \gamma (k_\gamma)+N(p_1)\to J/\psi (k_\psi)+N'(p_2) \ ,
\end{equation}
where we have used $J/\psi$ as an example, $k_\gamma$ and $k_\psi$ represent the momenta for incoming photon and outgoing $J/\psi$, $p_1$ and $p_2$ for incoming and outgoing nucleons. Similar diagrams have been considered in Ref.~\cite{Brodsky:2000zc} where it was argued that the three-gluon exchange diagrams dominate the near threshold production of $J/\psi$. However, from our analysis, the contribution from the three-gluon exchange diagrams vanishes due to $C$-parity conservation. Explicitly, the three gluons from the nucleon side carry symmetric color structure (such as $d_{abc}$)~\cite{Tong:2021ctu} while those from the heavy quarkonium ($J/\psi$) side are antisymmetric (such as $f_{abc}$). We notice that, however, $\eta_c$ production will be dominated by the three-gluon exchange diagrams.

In order to make the near threshold expansion more evident, it is useful to examine the relevant kinematics for the scattering amplitude. The center of mass energy squared and momentum transfer squared can be written as: $W_{\gamma p}^2=(k_\gamma+p_1)^2=(k_\psi+p_2)^2\sim M_V^2$ and $-t=-(p_2-p_1)^2\ll M_V^2$. In the heavy quark mass limit, we will have $p_1\cdot k_\gamma\sim p_1\cdot k_\psi\sim M_V^2$, whereas $p_2\cdot k_\gamma\sim p_2\cdot k_\psi\ll M_V^2$. In addition, the invariant mass of the $t$-channel two gluon is much smaller than heavy quarkonium mass. More importantly, the propagators in the heavy quark loop are all of the order $1/M_V$, e.g., $\left(k_1-k_\psi/2\right)^2-M_c^2=k_1^2-k_1\cdot k_\psi\sim -M_c^2$, where $M_c$ represents the Charm quark mass and $k_1$ for one of gluon momentum in the $t$-channel.

To compute the Feynman diagrams in Fig.~\ref{largetdiagram}, we follow the factorization argument for the hard exclusive processes~\cite{Lepage:1980fj}, where the leading contributions come from the three-valence quark Fock state of the nucleon. The three-quark states can be further classified into zero orbital angular momentum (OAM) and nonzero OAM components~\cite{Ji:2002xn}. We will first examine the contribution from zero OAM component, which is referred as twist-3 term. 

For the Fock state with zero OAM, the three quarks' momenta are at the same direction as the parenting nucleon and their total momentum equals to the nucleon momentum.  An important feature of this contribution is that the nucleon helicity is conserved. We take into account all permutations in the gluon attachments in both upper and lower parts of Fig.~\ref{largetdiagram}. The calculation is complicated but straightforward. In the end, we find that the scattering amplitude can be summarized as
\begin{align}
    {\cal A}_3 &=\langle J/\psi(\epsilon_\psi),N'_\uparrow|  \gamma(\epsilon_\gamma),N_\uparrow\rangle 
    \notag\\
    &=\int [d x][d y] \Phi(x_1,x_2,x_3)\Phi^*(y_1,y_2,y_3)\frac{1}{(-t)^3} \nonumber\\
    &~~\times {\cal M}_p(\{x_i\},\{y_i\}){\cal M}_\psi^{\mu\nu}(\epsilon_\gamma,\epsilon_{\psi},\{x_i\},\{y_i\})\nonumber\\
    &~~\times \bar U_\uparrow(p_2)\gamma^\mu U_\uparrow(p_1) \bar P^\nu\ ,\label{eq:twist3}
\end{align}
where $\bar P=(p_1+p_2)/2$, $\{x\}=(x_1,x_2,x_3 )$ represent the momentum fractions carried by the three quarks, $[d x]= d x_1 d x_2 d x_3\delta(1-x_1-x_2-x_3)$, and $ \Phi_3(x_i)$ is the twist-three distribution amplitude of the proton~\cite{Lepage:1980fj,Braun:1999te}. In the above equation, ${\cal M}_p$ and ${\cal M}_\psi^{\mu\nu}$ contains contributions from the nucleon and photon-quarkonium sides, respectively.  The spinor structure in Eq.~(\ref{eq:twist3}) is a consequence of the leading-twist amplitude which conserves the nucleon helicity. This is similar to the $A$ form factor calculation in Ref.~\cite{Tong:2021ctu}. Furthermore, we find that ${\cal M}_p$ can be simplified as
\begin{align}
{\cal M}_p= \frac{ C_B^2 }{96} (4\pi \alpha_s)^2 \left (2{\cal H}_3+{\cal H'}_3 \right),\quad  
\end{align}
where $C_B={2}/{3}$. The coefficient ${\cal H}_3$ can be summarized as
\begin{align}
{\cal H}_3=
 I_{13}+ I_{31}+ I_{23}+ I_{21}+ I_{12}+I_{32}
,
\end{align}
where $I_{ij}=\frac{\bar x_i+\bar y_i}{x_i x_j y_i y_j \bar{x}_i^2 \bar{y}_j^2}$ with $\bar x_i=1-x_i$, $\bar y_i=1-y_i$, and ${ \cal H}_3'={ \cal H}_3(y_1\leftrightarrow y_3)$.

The power behavior of $1/(-t)^3$ in Eq.~(\ref{eq:twist3}) comes from the propagators in the lower part and the $t$-channel gluons. This behavior is also consistent with the conventional power counting analysis~\cite{Brodsky:1973kr,Matveev:1973ra}. However, the final result for the differential cross section will depend on the amplitude squared in the threshold limit $\chi\to 1$. For that, we find, 
\begin{equation}
    |\overline{{\cal A}_3}|^2=(1-\chi)G_\psi G_{p3}(t)G_{p3}^*(t)  \ ,\label{eq:twist3p}
\end{equation}
which actually vanishes at the threshold. In the above, $G_\psi$ is defined as,
\begin{equation}
 G_\psi   =C_N^2\frac{ 32 \pi^2 \alpha e_c^2(4\pi \alpha_s)^6}{3 M_\psi ^3} \langle   {\cal O}_1^{\psi}({}^3S_1) \rangle\ ,
\end{equation}
where $C_N=\frac{2}{27}$, $\langle  {\cal O}_1^\psi({}^3S_1) \rangle$ is the color-singlet NRQCD matrix element for $J/\psi$. $G_p$ follows the form factor factorization and can be written as
\begin{equation}
    G_{p3}(t)=\frac{1}{t^2}\int [dx][dy]\Phi_3(\{x\})\Phi_3^*(\{y\})\left[2{\cal T}_3+{\cal T}_3'\right] \ ,
\end{equation}
where ${\cal T}_3'={\cal T}_3(y_1\leftrightarrow y_3)$ and ${\cal T}_3$ can be obtained from ${\cal H}_3$ by the replacement $I_{ij}\rightarrow \frac{\bar x_i+\bar y_i}{\bar x_i}I_{ij}$ .
Combining $G_{p3}$ and $G_{p3}^*$, this leads to $1/(-t)^4$ power behavior for the amplitude squared, which is consistent with the conventional power counting analysis. However, this contribution is suppressed at the threshold.

The suppression factor $(1-\chi)$ comes from the spinor structure in Eq.~(\ref{eq:twist3}). Similar suppression has also been found in the generalized parton distribution (GPD) framework~\cite{Ji:1996ek,Ji:1996nm,Mueller:1998fv,Radyushkin:1996nd,Collins:1996fb,Diehl:2003ny,Belitsky:2005qn} for exclusive photoproduction of $J/\psi$, where the contribution from $H_g$ GPD is associated with a factor of $(1-\xi)$ with $\xi$ being the skewness parameter~\cite{Hoodbhoy:1996zg}, see also, \cite{Ivanov:2004vd,Koempel:2011rc} and references therein.

In order to obtain a nonvanishing contribution at the threshold, we have to go beyond the leading-twist contributions, such as those from three-quark Fock state with nonzero OAM. In the following, we consider the three-quark Fock state with one unit OAM~\cite{Ji:2002xn}. We call this as twist-4 contribution because it depends on the twist-4 distribution amplitudes.

Two important features emerge for nonzero OAM contributions. First, as shown in Fig.~\ref{largetdiagram}, the partonic scattering amplitudes conserve the quark helicities. However, because of a nonzero OAM for one of the three-quark state, the helicity of the nucleon states will be different. In the sense that nonzero quark OAM contributes to the hadron helicity-flip amplitude. Second, in order to get a nonzero contribution, we have to perform the intrinsic transverse momentum expansion for the hard partonic scattering amplitudes~\cite{Belitsky:2002kj}, which will introduce an additional suppression factor of $1/(-t)$. Since one unit OAM is involved in the calculation, the linear term in this expansion contributes to the final result, that can be written in terms of twist-four distribution amplitude of the nucleon~\cite{Braun:1999te,Belitsky:2002kj}. Here we summarize the final expression with the power counting result, 
\begin{eqnarray}
      {\cal A}_4&=&\langle J/\psi(\epsilon_\psi),N'_\uparrow|  \gamma(\epsilon_\gamma),N_\downarrow\rangle 
    \notag\\
&=&\int [dx][dy]\Psi_4(\{x\})\Phi_3^*(\{y\}){\cal M}_\psi^{(4)}\left(\{x\},\{y\}\right)\nonumber\\
&&\times \bar U_\uparrow(p_2) U_\downarrow(p_1) \frac{M_p}{(-t)^3}\ ,\label{eq:twist4}
\end{eqnarray}
where $\Psi_4$ is one of the twist-four distribution amplitudes of the proton related to the three quark Fock state with one unit OAM~\cite{Braun:2000kw,Belitsky:2002kj}. Similar contribution can be obtained for another twist-four distribution amplitude $\Phi_4$. Here we emphasize a couple of important points. First, the factor $M_p$ in Eq.~(\ref{eq:twist4}) indicates it is a higher-twist effect. Explicitly, it comes from the parameterization of the twist-four distribution amplitude~\cite{Braun:1999te}. Second, the nucleon helicity-flip is manifest in the spinor structure. This amplitude is negligible at high energy, but will be important at the threshold, because it is not suppressed in the limit of $\chi\to 1$. The amplitude squared can be written as
\begin{eqnarray}
    |\overline{{\cal A}_4}|^2=\widetilde m_t^2 G_\psi G_{p4}(t) G_{p4}^*(t) \ , \label{eq:twist4p}
\end{eqnarray}
where $\widetilde m_t^2=M_p^2/(-t)$, $G_\psi$ is the same as above, and $G_{p4}$ depend on the twist-three and twist-four distribution amplitudes,
\begin{eqnarray}
    G_{p4}(t)=\frac{1}{t^2}\int [dx][dy]\Psi_4(\{x\})\Phi_3^*(\{y\}){\cal T}_{\Psi 4} \ .
\end{eqnarray}
The coefficient ${\cal H}_{\Psi 4}$ is much more complicated as compared to ${\cal H}_3$. 
 
Eqs.~(\ref{eq:twist4p}) and (\ref{eq:twist3p}) are the most important results of our analysis. Comparing these two, we find that the twist-four contribution is suppressed in $1/t$ but enhanced at the threshold. These two features can be used to disentangle their contributions in experiments. If we limit our discussions in the threshold region, the only contribution comes from the twist-four term. 

In the literature, the near threshold heavy quarkonium production amplitude has been written in terms of the gluonic form factors. The gluonic form factors at large $(-t)$ have been recently calculated in Ref.~\cite{Tong:2021ctu}. Their results show that $A_g$ form factor is similar to the above helicity-conserved amplitude, whereas $B_g$ and $C_g$ form factors are associated with the helicity-flip amplitudes. By comparing this to the above results, we conclude that the $A_g$ form factor will not be responsible for heavy quarkonium production at the threshold. Now, the question becomes: can we re-write the near threshold helicity-flip amplitude (Eq.~(\ref{eq:twist4})) in terms of the gravitation form factors $B_g/C_g$ or a combination (including $\langle N'|F^2|N\rangle$)? From a detailed comparison, unfortunately, we are not able to build a direct connection between them~\footnote{We have also checked this for a simpler process such as $\gamma \pi\to J/\psi \pi$ and found no connection to the gluonic gravitational form factor of pion at large $(-t)$, which have also been calculated in Ref.~\cite{Tong:2021ctu}.}. This may impose a challenge to interpret the near threshold heavy quarkonium photoproduction as a measure to the gluonic gravitational form factors. It could well be that this interpretation only breaks down at large $(-t)$. Therefore, although it is a disappointing answer, the question itself deserves further investigations. 

Recently, Ref.~\cite{Guo:2021ibg} has suggested that the GPD formalism could be applied in near threshold heavy quarkonium production, see also the discussion in Ref.~\cite{Hatta:2021can}. It will be interesting to check this statement with our results, where the gluon GPDs at large momentum transfer can be calculated following the example of the quark GPDs in~\cite{Hoodbhoy:2003uu}.

\section{Phenomenology Applications}

To summarize the results in the previous section, we obtain the differential cross section at large $(-t)$
\begin{eqnarray}
\frac{d\sigma}{dt}|_{(-t)\gg\Lambda_{QCD}^2}&=&\frac{1}{16\pi(W_{\gamma p}^2-M_p^2)^2}\left( |\overline{{\cal A}_3}|^2+|\overline{{\cal A}_4}|^2\right)\nonumber\\
&\approx &\frac{1}{(-t)^4}\left[(1-\chi){\cal N}_3+\widetilde m_t^2 {\cal N}_4\right]\ ,\label{eq:diff}
\end{eqnarray}
where ${\cal N}_3$ and ${\cal N}_4$ represent the twist-three and twist-four contributions, respectively. They depend on the associated distribution amplitudes of the nucleon. We note that there is no interference between these two, because their helicity configurations are different. 

\begin{figure}[tpb]
\includegraphics[width=0.49\columnwidth]{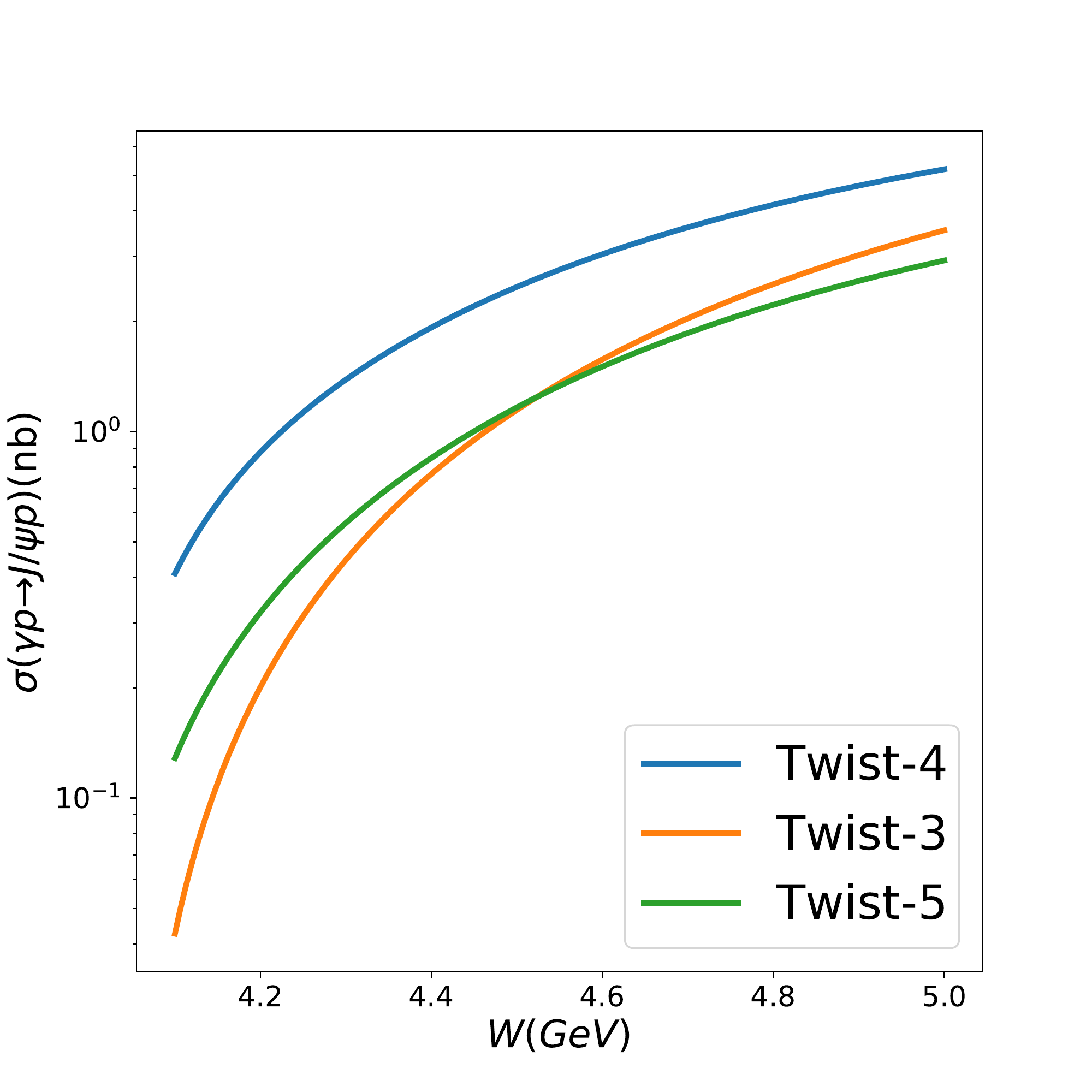}
\includegraphics[width=0.49\columnwidth]{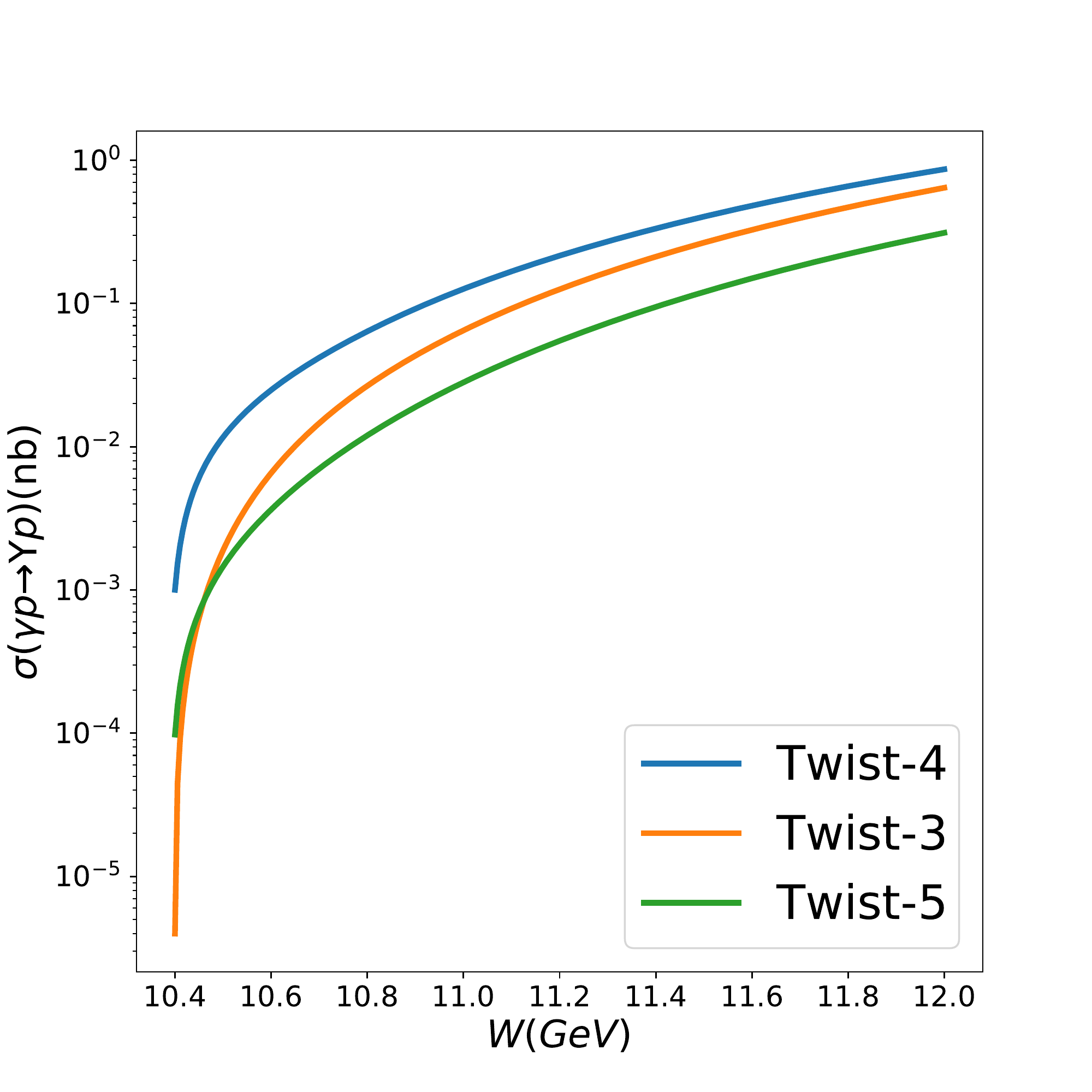}
    \caption{Parametric comparison between the two contributions (twist-3 and twist-4) to the near threshold heavy quarkonium production as functions of the center mass energy $W_{\gamma p}$ for $J/\psi$ (left) and $\Upsilon$ (right). Here we plot the cross section contributions in arbitrary unit, assuming the same coefficients for ${\cal N}_3$ and ${\cal N}_4$ in Eq.~(\ref{eq:diff}). A potential twist-5 contribution is also shown in these plots.}
    \label{fig:twist}
\end{figure}

The above two contributions have different power behavior for the differential cross sections, one with $1/(-t)^4$ and one with $1/(-t)^5$. Although the current experimental data can not distinguish them, high precision future experiments~\cite{Chen:2014psa,AbdulKhalek:2021gbh,Joosten:2018gyo} will be able to provide an important test. The most important consequence of our power counting analysis is that the leading-twist contribution is suppressed at the threshold. Away from the threshold point, it will start to contribute and may dominate at large $(-t)$ because of the leading power feature.

This will be reflected in the total cross section contributions as well, for that we have
\begin{equation}
    \sigma(W_{\gamma p})=\int_{t_{min}}^{t_{max}} \frac{d\sigma}{dt} (t)\ ,
\end{equation}
where $t_{min}$ and $t_{max}$ depend on the center of mass energy $W_{\gamma p}$. At the threshold point we have $t_{min}=t_{max}$, so that the total cross section vanishes. In Fig.~\ref{fig:twist}, we compare the above two contributions as functions of $W_{\gamma p}$ for $J/\psi$ and $\Upsilon$, respectively, assuming ${\cal N}_3={\cal N}_4$ for an order of magnitude estimate. In order to smooth the contributions at small-$(-t)$, we modify the above $t$ by $t-\Lambda^2$ where $\Lambda=1~{\rm GeV}$ to represent a non-perturbative scale. We note that the dominance of twist-4 contribution in both $J/\psi$ and $\Upsilon$ productions around the threshold is insensitive to the choice of $\Lambda$.   

We emphasize that the higher-twist contribution beyond the twist-four will be negligible. As an example, in Fig.~\ref{fig:twist} we plot a contribution from a potential twist-5 term which scales as $1/(-t)^6$ for the differential cross section. It is similar to ${\cal N}_4$ term in Eq.~(\ref{eq:diff}) as $\widetilde{m}_t^4 {\cal N}_5$ and we assume ${\cal N}_5={\cal N}_4={\cal N}_3$ in these curves. From these plots, we conclude that the twist-four term is the dominant contribution for near threshold photoproduction.

\begin{figure}[tpb]
\includegraphics[width=0.49\columnwidth]{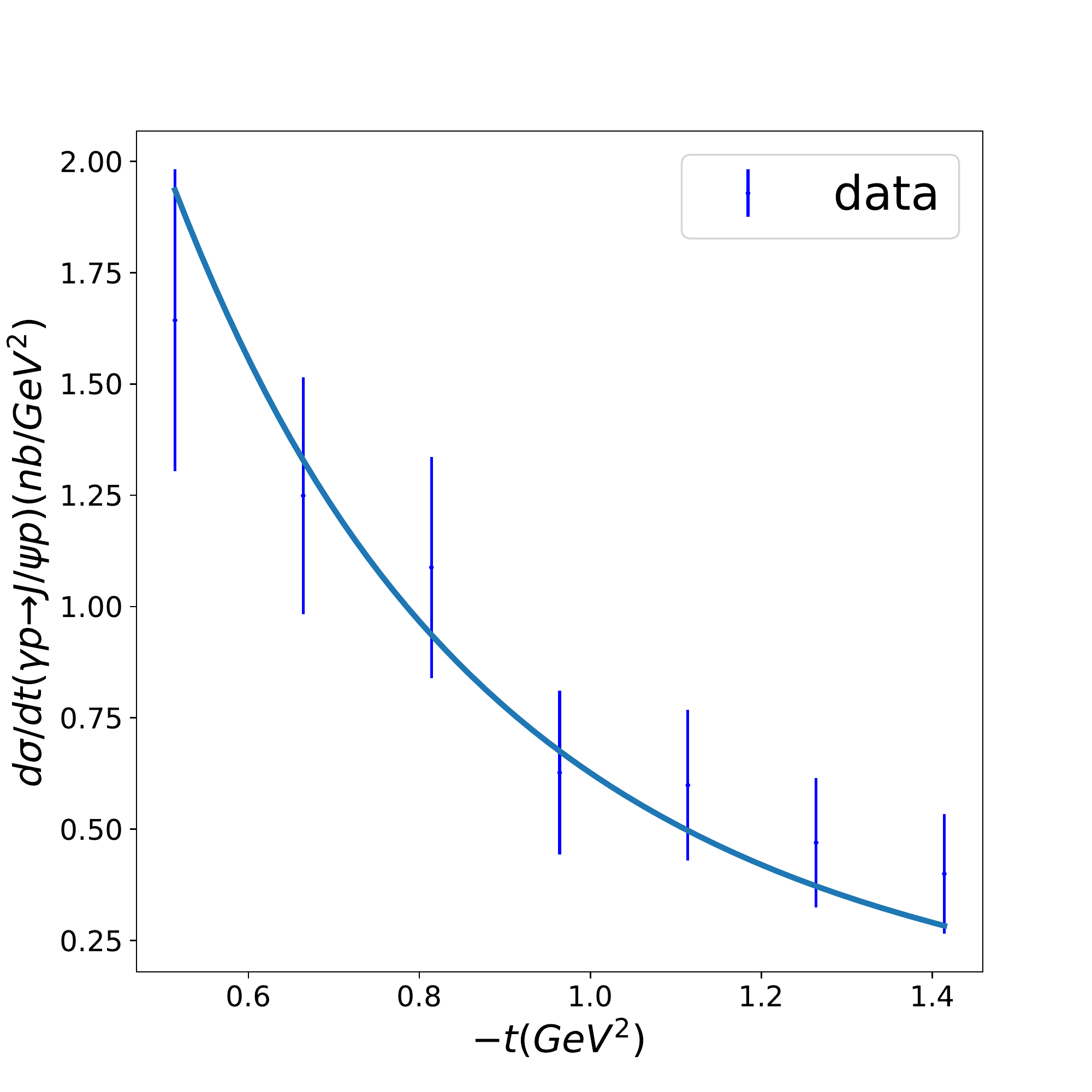}
\includegraphics[width=0.49\columnwidth]{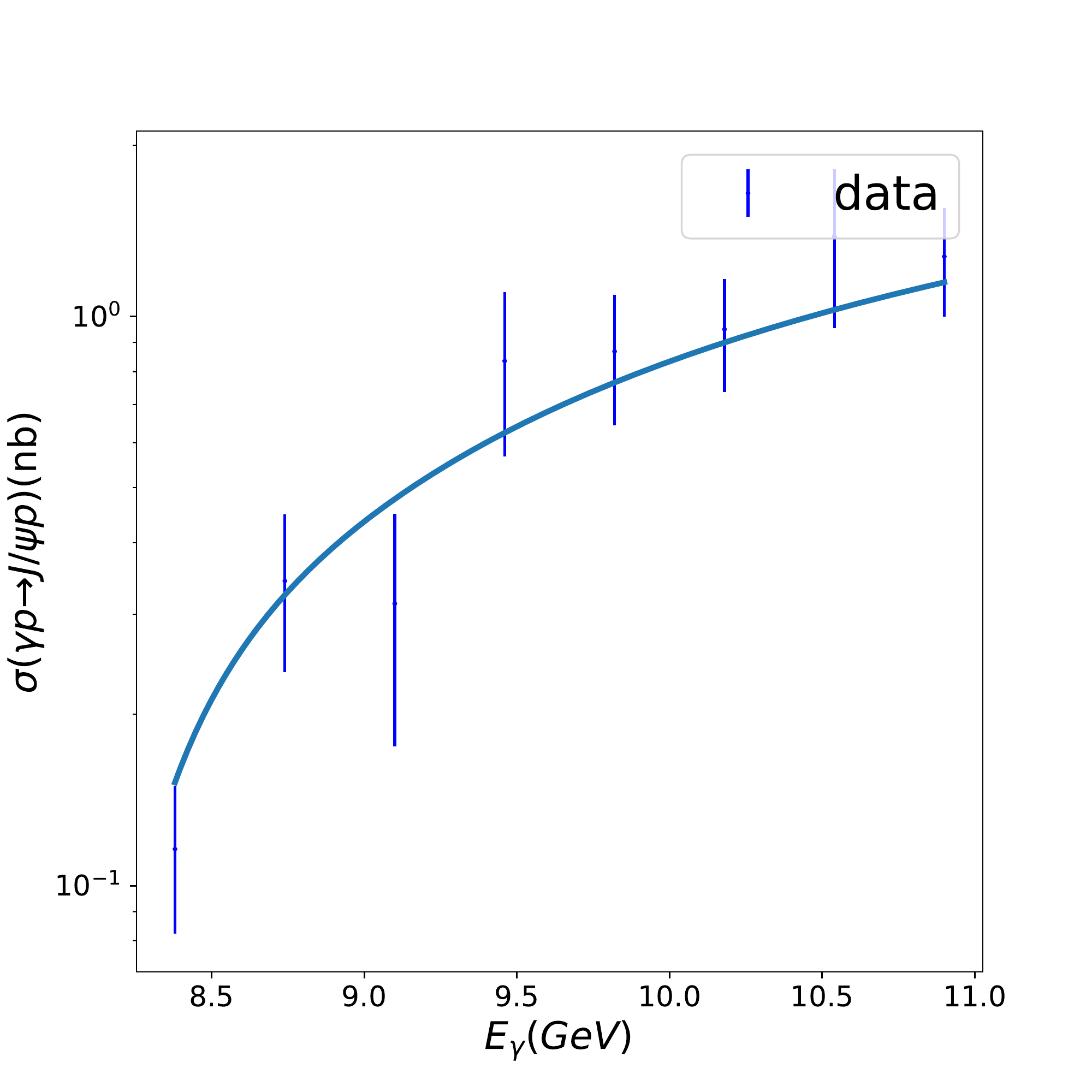}
    \caption{Fit the experimental data from the GlueX collaboration~\cite{Ali:2019lzf} with the leading contribution from the twist-four term in the differential cross section: (left) the differential cross section at $E_\gamma\approx 10.72~{\rm GeV}$ assuming $\langle t_{min}\rangle\approx -0.44 {\rm GeV}^2$; (right) the total cross section near the threshold.}
    \label{fig:fitgluex}
\end{figure}

In Fig.~\ref{fig:fitgluex}, we apply our power counting analysis to the experimental data from the GlueX collaboration~\cite{Ali:2019lzf}. For the illustration purpose, we only include the twist-four term in the differential cross section,
\begin{eqnarray}
\frac{d\sigma}{dt}|^{twist-4}=\frac{N_0}{(-t-\Lambda^2)^5}\ ,
\end{eqnarray}
and the total cross section is calculated by integrating over $t$. We fit the GlueX data with two parameters $N_0$ and $\Lambda$,
\begin{equation}
    \Lambda^2=1.41\pm 0.20~{\rm GeV}^2\ ,~~~N_0=51\pm 22 ~{\rm nb * GeV^8} \ ,
\end{equation}
with a $\chi^2/d.o.f.=0.48$. Fig.~\ref{fig:fitgluex} shows that our predictions are consistent with the experimental data. The comparison also shows that there may need further improvement by including subleading contributions when the energy is away from the threshold. In the above analysis we only take into account the power counting predictions. It will be interesting to compute the differential cross sections with the nucleon distribution amplitudes~\cite{Braun:2000kw}. We will carry out a comprehensive study in the future.

It is important to note that the above power counting analysis was derived for large $(-t)$ differential cross sections. The consistency between our predictions and the GlueX data shall encourage further theoretical developments, in particular, in the lower momentum transfer region where one can study the interplay between the perturbative and non-perturbative physics. Regarding this point, the comparison between $J/\psi$ and $\Upsilon$ productions will play an important role, because they offer different kinematic coverage of momentum transfer due to their large mass difference. We expect these processes will be extensively investigated at the future EIC~\cite{AbdulKhalek:2021gbh,Anderle:2021wcy}.

\section{Summary}

In this paper, we have performed a perturbative QCD analysis for the near threshold heavy quarkonium photoproduction at large momentum transfer. We have shown that the so-called three-gluon exchange diagrams do not contribute. The contribution from the leading Fock state with zero OAM of nucleon is suppressed at threshold. The differential cross section is dominated by the contribution from nonzero OAM Fock state and has a power behavior of $1/(-t)^5$. This prediction is different from previous assumptions in the literature. We have applied the power counting result to the GlueX data and found that they agree with each other.

Through explicit calculations, we have shown that there is no direct connection between the near threshold heavy quarkonium production and the gluonic gravitational form factors of the proton. We note that, however, under certain approximations the connection between them can be built through a GPD formalism~\cite{Hatta:2021can,Guo:2021ibg}.  

{\bf Acknowledgments:} We thank Xiangdong Ji for communications on their preprint of Ref.~\cite{Guo:2021ibg}. We thank Stan Brodsky, Yoshitaka Hatta, Xiangdong Ji, Igor Strakovsky, Nu Xu for discussions and comments. This material is based upon work supported by the U.S. Department of Energy, Office of Science, Office of Nuclear Physics, under contract numbers DE-AC02-05CH11231. 


\end{document}